\documentclass[11pt,uplatex]{article}
\usepackage{graphicx}
\usepackage{color}
\usepackage{amsmath, amssymb, amsthm, ascmac, caption, comment, enumerate, ascmac, bm}
\usepackage{latexsym, mathabx, mathrsfs, mathtools, psfrag, setspace, subcaption, booktabs, soul}
\usepackage{natbib}
\definecolor{mycolor}{cmyk}{0.80, 0.20, 0.25, 0}
\usepackage[colorlinks=true, linkcolor=mycolor, filecolor=mycolor, urlcolor=mycolor, citecolor=mycolor, driverfallback=dvipdfmx]{hyperref}
\usepackage[margin = 2.2cm]{geometry}

\usepackage{algpseudocode}
\usepackage{algorithm}
\floatname{algorithm}{Procedure}

\usepackage{titlesec}
\titleformat*{\section}{\Large\bfseries\sffamily}
\titleformat*{\subsection}{\large\bfseries\sffamily}
\titleformat*{\subsubsection}{\large\bfseries\sffamily}

\renewcommand{\hat}{\widehat}
\renewcommand{\tilde}{\widetilde}

\numberwithin{equation}{section}
 
\allowdisplaybreaks[2]

\theoremstyle{definition}
\newtheorem{theorem}{Theorem}[section]

\newtheorem{definition}{Definition}[section]

\newtheorem{example}{Example}[section]
\AtBeginEnvironment{example}{
  \pushQED{\qed}
}
\AtEndEnvironment{example}{\popQED\endexample}

\newtheorem{remark}{Remark}[section]
\AtBeginEnvironment{remark}{
  \pushQED{\qed}
}
\AtEndEnvironment{remark}{\popQED\endremark}

\title{Randomization Test for the Specification of Interference Structure}

\author{Tadao Hoshino\thanks{School of Political Science and Economics, Waseda University, 1-6-1 Nishi-waseda, Shinjuku-ku, Tokyo 169-8050, Japan. Email: \href{mailto:thoshino@waseda.jp}{thoshino@waseda.jp}.} \
and Takahide Yanagi\thanks{Graduate School of Economics, Kyoto University, Yoshida Honmachi, Sakyo, Kyoto 606-8501, Japan. Email: \href{mailto:yanagi@econ.kyoto-u.ac.jp}{yanagi@econ.kyoto-u.ac.jp}}}

\date{December 2023}

\begin{document}

\maketitle

\begin{abstract}
    This study considers testing the specification of spillover effects in causal inference.
    We focus on experimental settings in which the treatment assignment mechanism is known to researchers.
    We develop a new randomization test utilizing a hierarchical relationship between different exposures.
    Compared with existing approaches, our approach is essentially applicable to any null exposure specifications and produces powerful test statistics without a priori knowledge of the true interference structure.
    As empirical illustrations, we revisit two existing social network experiments: one on farmers' insurance adoption and the other on anti-conflict education programs.
    
    \bigskip
	
	
	\bigskip 
	
\end{abstract}

\clearpage

\section{Introduction}

Causal inference in the presence of cross-unit treatment interference has gained increasing attention in the literature.
Previous studies have highlighted the importance of accounting for potential treatment spillovers through empirical applications in many fields, including economics, education, epidemiology, and political science.
Because individuals generally have different interaction networks, it is typically impossible to identify any meaningful causal parameters without some simplifying assumptions on the interference structure (\citealp{imbens2015causal}).
A popular approach in the literature is to introduce some \textit{exposure mapping} that summarizes the impacts from other individuals' treatments into lower-dimensional statistics (\citealp{aronow2021spillover}).
For example, \cite{hong2006evaluating} studied the impact of school retention on later academic performance, assuming that other students' retention may affect their own performance depending only on whether their school has a higher or lower retention rate.
As another example, \cite{leung2020treatment} considered a treatment spillover model in which other individuals' treatments affect one's potential outcome only through the number of treated neighbors and the total size of the neighbors.

For the estimated spillover effects to be meaningful, we must justify the specification of the exposure mapping.
However, there is no general theoretical guidance on what exposure mapping should be used, and if the chosen specification is inappropriate, the resulting causal inference may be misleading (e.g., failure to detect treatment spillovers).\footnote{
    Some recent studies have investigated under what conditions one can estimate the meaningful causal parameters even when the exposure mapping is misspecified or not explicitly specified (\citealp{aronow2017estimating}; \citealp{savje2021average}; \citealp{leung2022causal}; \citealp{hoshino2023causal}).
	A common finding in these studies is that only if the network dependence is sufficiently weak can we identify some composite causal parameters.
	Nevertheless, for a general form of network interference, knowledge of the true exposure is still essential.
}
An approach to directly address this issue is to statistically test the specification, where the null hypothesis of interest is whether a given exposure mapping is a ``correct'' choice (a more formal argument will be given later).
This is the aim of the present study.
Specifically, with a focus on experimental situations where the treatment assignment mechanism is known to researchers, we develop new randomization tests for testing the specification of general exposure mappings.

Unlike in the standard Fisher randomization test, the null hypothesis of interest here is not ``sharp'' in general in the sense that only a subset of the potential outcomes are imputable from the observed outcomes. 
Consequently, to perform randomization tests, we must carefully select the appropriate subsets of units and treatment assignments, which we call the \textit{focal subpopulation} and \textit{focal assignments}, respectively.
We are not the first to consider this type of ``conditional'' randomization test.
In the literature, \cite{aronow2012general}, \cite{athey2018exact}, \cite{basse2019randomization}, and \cite{puelz2022graph} have considered similar conditional randomization tests for testing the structure of treatment spillovers (they did not characterize their methods as a ``specification test'' though).
However, they did not discuss a unified framework for constructing appropriate test statistics for general exposure mappings; instead, they focused on several specific nulls or specific experimental setups.

The main contributions of this study are three-fold:
First, introducing the notion of \textit{coarseness} of exposure mappings, we propose a novel randomization testing approach that can test virtually any null exposures and automatically produce model-free test statistics equipped with reasonable power in most situations.
Second, we prove the validity of our testing procedure under relatively mild conditions.
Lastly, we apply our specification test of exposure mappings to two prominent social network experiments in the literature: one on farmers' insurance adoption by \cite{cai2015social} and the other on anti-conflict intervention school programs by \cite{paluck2016changing}.
In both applications, our tests provide statistical evidence for the existence of spillover effects and, in particular, suggest that having at least one treated peer may serve as a good summary statistic for the treatment spillovers.

The rest of this paper is organized as follows.
Section \ref{sec:test} introduces our randomization test.
In Section \ref{sec:MC}, we report the results of the Monte Carlo experiments.
Section \ref{sec:empiric} presents two empirical case studies. Finally, Section \ref{sec:conclusion} concludes the paper.
The accompanying {\ttfamily R} package \href{https://tkhdyanagi.github.io/testinterference/}{{\ttfamily testinterference}} is available from the authors' websites.

\section{Randomization Test} \label{sec:test}

\subsection{Setup}

Consider a finite population of size $n$. 
Each unit is indexed by $i \in [n]$, where, for a positive integer $a$, we denote $[a] = \{1, \ldots, a\}$.
Throughout this paper, we focus on the experimental setup in which an experimenter assigns binary treatments $\bm{Z} = (Z_1, \dots, Z_n) \in \mathcal{Z}$ to $n$ units according to a known assignment probability $\mathbb{P}_{\bm{Z}}$.
Here, $\mathcal{Z} \coloneqq \{ \bm{z} \in \{ 0, 1 \}^n : \mathbb{P}_{\bm{Z}}(\bm{z}) > 0\}$ denotes the set of all possible assignment patterns.

In this section, we mainly consider the case wherein all units fully comply with their assigned treatments.
Thus, for all $i$, we do not distinguish between the assigned treatment $Z_i$ and the actual treatment take-up of $i$, which we denote as $D_i$.
The case where noncompliance is allowed such that $Z_i \neq D_i$ for some units will be discussed in Subsection \ref{subsec:imperfect}.

The outcome variable is $Y_i \in \mathcal{Y} \subseteq \mathbb{R}$.
In the most general treatment spillover model, each $Y_i$ may be affected by all elements of $\bm{Z}$.
Denoting as $Y_i(\bm{z})$ the potential outcome when $\bm{Z} = \bm{z}$, we have $Y_i \equiv Y_i(\bm{Z})$.
We assume that the potential outcomes are non-stochastic and that any random variation arises only from the randomness of the treatment assignment (i.e., the \textit{design-based} approach).\footnote{
    Alternatively, one may assume that the potential outcomes are random and view the analysis as being conditioned on all potential outcomes.
    }

Let $E: [n] \times \mathcal{Z} \to \mathcal{E}$ denote an exposure mapping, where $\mathcal{E}$ is a finite set.
For notational simplicity, we often write $E_i(\bm{z}) \equiv E(i, \bm{z})$, whose range is $\mathcal{E}_i$.
By construction, $\mathcal{E}_i \subseteq \mathcal{E}$.
In general, $\mathcal{E}$ may depend on $n$ and it is possible that $E_i(\bm{z})$ and $E_j(\bm{z})$ have different functional forms.

\begin{definition}[Correct exposure mapping]
    An exposure mapping $E$ is correct if the potential outcome value is uniquely determined by $E$; that is, for all $i \in [n]$ and $\bm{z}, \bm{z}' \in \mathcal{Z}$, $E_i(\bm{z}) = E_i(\bm{z}')$ $\Longrightarrow$ $Y_i(\bm{z}) = Y_i(\bm{z}')$.
\end{definition}

A correct exposure mapping always exists and is not necessarily unique.
In an extreme case, the identity mapping $E_i(\bm{z}) = \bm{z}$ is always correct.
In another example, when the stable unit treatment value assumption is fulfilled, $E_i(\bm{z}) = z_i$ and $E_i(\bm{z}) = (z_i, z_j)$ for any $j$ are both correct specifications.
Fisher's sharp null of no treatment effect can also be viewed as a special case where the exposure mapping $E_i(\bm{z})$ is a constant function independent of $\bm{z}$.

For an exposure mapping $E$ and $\bm{z} \in \mathcal{Z}$, we define the level set $\mathcal{L}_i(\bm{z} \mid E) \coloneqq \{ \bm{z}' \in \mathcal{Z} : E_i(\bm{z}') = E_i(\bm{z}) \}$.
If $E$ is correct, then all treatment assignments in $\mathcal{L}_i(\bm{z} \mid E)$ lead to the same potential outcome value for $i$.
In other words, when $E$ is correct, we can define a corresponding potential outcome function $y_i: \mathcal{E}_i \to \mathcal{Y}$ such that
\begin{align*}
    y_i(E_i(\bm{z})) = Y_i(\bm{z}') \quad \text{for all $\bm{z}' \in \mathcal{L}_i(\bm{z} \mid E)$}.
\end{align*}
In particular, $Y_i = y_i(E_i(\bm{Z}))$ holds under a correct $E$.

Once an exposure mapping is given, it induces a partition of $\mathcal{Z}$ specific to each $i$, where each block is given by the corresponding $\mathcal{L}_i(\bm{z} \mid E)$.
In particular, when $E$ is correct, $\mathcal{L}_i(\bm{z} \mid E)$ provides a partition by the equivalence class in terms of the potential outcome value. 
Based on the coarseness of the partition, we define the coarseness of the exposure mappings as follows:

\begin{definition}[Coarseness]\label{def:coarseness}
    For exposure mappings $E: [n] \times \mathcal{Z} \to \mathcal{E}$ and $E': [n] \times \mathcal{Z} \to \mathcal{E}'$, $E$ is coarser than $E'$ if there is a surjective mapping $c: \mathcal{E}' \to \mathcal{E}$ such that $c(E'(i, \bm{z}')) = E(i, \bm{z})$ for all $i \in [n]$, $\bm{z} \in \mathcal{Z}$, and $\bm{z}' \in \mathcal{L}_i(\bm{z} \mid E)$.
    When this holds true, we write $E' \supset E$.
\end{definition}

The definition says that $E$ is coarser than $E'$ if the value of $E$ is pinned down once that of $E'$ is given.
A concept similar to Definition \ref{def:coarseness} can be found in \cite{vazquez2023identification}.
The dimensions of $\mathcal{E}$ and $\mathcal{E}'$ are generally different.
For example, when the data are composed of $n/2$ pairs, we may consider $E_i(\bm{z}) = z_i$ and $E_i'(\bm{z}) = (z_i, z_{i'})$, where $i'$ denotes $i$'s partner. 
By definition, the finest exposure mapping is the identity mapping $E_i(\bm{z}) = \bm{z}$ and the coarsest one is $E_i(\bm{z}) = a$ for some constant $a$ independent of $\bm{z}$, which corresponds to the case of no treatment effect whatsoever.

\subsection{Testing procedure}

We would like to test whether an exposure mapping $E^0:[n] \times \mathcal{Z} \to \mathcal{E}^0$ is correct:
\begin{align*}
    \mathbb{H}_0: \; \text{$E^0$ is a correct exposure mapping.}
\end{align*}
Let $E^1:[n] \times \mathcal{Z} \to \mathcal{E}^1$ denote another exposure mapping such that $E^1 \supset E^0$.
Then, there exists a mapping $c^{1 \to 0}: \mathcal{E}^1 \to \mathcal{E}^0$ that satisfies $c^{1 \to 0}(E_i^1(\bm{z}')) = E_i^0(\bm{z})$ for $\bm{z}' \in \mathcal{L}_i(\bm{z} \mid E^0)$.
If $\mathbb{H}_0$ is true, both $E^0$ and $E^1$ are correct, ensuring the existence of potential outcome functions $y_i^0: \mathcal{E}_i^0 \to \mathcal{Y}$ and $y_i^1: \mathcal{E}_i^1 \to \mathcal{Y}$ satisfying $Y_i = y_i^0(E_i^0(\bm{Z})) = y_i^1(E_i^1(\bm{Z}))$.
We define
\begin{align*}
    \tilde{\mathcal{E}}^1_i \equiv \tilde{\mathcal{E}}^1_i(\bm{Z}) \coloneqq \{e^1 \in \mathcal{E}_i^1 : c^{1 \to 0}(e^1) = E^0_i(\bm{Z}) \},
\end{align*}
namely, the set of $E^1_i$ values that map to $E^0_i(\bm{Z})$ through $c^{1\to 0}$.
Then, under $\mathbb{H}_0$,
\begin{align*} 
        Y_i 
        = y_i^0(E^0_i(\bm{Z}))
        = y_i^1(e^1)
        \quad \text{for all $e^1 \in \tilde{\mathcal{E}}^1_i$}.
\end{align*}
Thus, the values of all $\{y_i^1(e^1) : e^1 \in \tilde{\mathcal{E}}^1_i\}$ are identically imputable as $Y_i$.
By construction, $E^1_i(\bm{Z}) \in \tilde{\mathcal{E}}^1_i$ is always satisfied, implying that $|\tilde{\mathcal{E}}^1_i| \ge 1$ uniformly in $i$, where $|\mathcal{A}|$ denotes the cardinality of a set $\mathcal{A}$.

Our key idea is to test whether the following equality is true for all focal units:
\begin{align} \label{eq:nullequality}
    y_i^1(e^1_j) = y_i^1(e^1_k)
    \quad
    \text{for all $e^1_j, e^1_k \in \tilde{\mathcal{E}}^1_i$}.
\end{align}
To be more specific, for some $\kappa \ge 2$, let $\mathcal{N}(\kappa) \coloneqq \{i \in [n] : |\tilde{\mathcal{E}}^1_i| = \kappa\}$ denote the set of units with $\kappa$ realizations of $E^1$ that are consistent with the actual $E^0$ value.
Then, we choose a set of focal units as $\mathcal{S} \subseteq \mathcal{N}(\kappa)$, the \textit{focal subpopulation}.
How to construct $\mathcal{S}$ in practice will be discussed later.
Although the construction of $\mathcal{S}$ can be stochastic in general, the following analysis treats $\mathcal{S}$ as given.

For a focal subpopulation $\mathcal{S}$, we define the set of \textit{focal assignments} as
\begin{align*}
    \mathcal{C}^\mathcal{S} 
    \equiv \mathcal{C}^\mathcal{S}(\bm{Z})
    \coloneqq \{ \bm{z} \in \mathcal{Z} : E^1_i(\bm{z}) \in \tilde{\mathcal{E}}^1_i \;\; \text{for all $i \in \mathcal{S}$} \},
\end{align*}
which is the set of treatment assignments inducing only $E^1$ values that match the observed $E^0$ values for all members in $\mathcal{S}$. 
As long as $\bm{z}$'s are taken from $\mathcal{C}^\mathcal{S}$, for any $i \in \mathcal{S}$, it is satisfied that $E_i^0(\bm{z}) = E_i^0(\bm{Z})$ for all such $\bm{z}$, whereas we can generate variations in $E_i^1$ values within $\tilde{\mathcal{E}}^1_i$.
Our randomization test computes the null distribution of a test statistic by randomly sampling the assignments $\bm{z}$'s from $\mathcal{C}^\mathcal{S}$ and checking whether \eqref{eq:nullequality} holds true for all $i \in \mathcal{S}$.
In practice, when $\mathcal{C}^\mathcal{S}$ is too vast to compute, one may impose additional conditions on $\mathcal{C}^\mathcal{S}$ to reduce its size, which might reduce the power of the test, but does not lose its validity. 

We generally cannot use the entire $\mathcal{N}(\kappa)$ as the focal subpopulation, but need to form $\mathcal{S}$ as a subset of $\mathcal{N}(\kappa)$ to retain sufficient variations in the focal assignments in $\mathcal{C}^\mathcal{S}$.
The following example would be helpful in understanding this.

\begin{example}\label{ex:pair}
    Suppose that the population is composed of $n/2$ couples.
    Let $E^0_i(\bm{z}) = z_i$ and $E^1_i(\bm{z}) = (z_i, z_{i'})$, where $i'$ indicates $i$'s partner.
    Trivially, $E^1 \supset E^0$ with $c^{1 \to 0}$ being a function that selects the first element of $E^1_i(\bm{z})$.
    If $E^0$ is a correct exposure, so is $E^1$, implying that
    \begin{align*}
        Y_i = y^0_i(Z_i) = y^1_i(e^1) \;\; \text{for all $e^1 \in \tilde{\mathcal{E}}^1_i = \{(Z_i,0), (Z_i, 1)\}$ and $i \in [n]$.}
    \end{align*}
    Thus, $\mathcal{N}(2)$ coincides with the entire population $[n]$ if all individuals are treatment eligible.
    If we use the entire $\mathcal{N}(2)$ as the focal subpopulation $\mathcal{S}$, we must shuffle the treatment assignments while keeping $(z_i, z_{i'}) = (Z_i, Z_{i'})$ for all pairs (otherwise, the potential outcomes are not imputable for both partners in each pair).
    However, $\bm{Z}$ is clearly the only treatment assignment that satisfies such a constraint and the randomization test is infeasible.
    In this example, the most reasonable focal subpopulation would be obtained by randomly selecting one unit from each pair, such that $|\mathcal{S}| = n/2$.
    The corresponding set of focal assignments is $\mathcal{C}^\mathcal{S} = \{ \bm{z} \in \mathcal{Z} : z_i = Z_i \;\; \text{for all $i \in \mathcal{S}$}\}$.
\end{example}

As shown in the next example, our framework encompasses the Fisher randomization test of no treatment effect as a special case.

\begin{example}
    Let $E^0_i(\bm{z}) = a$, where $a$ is independent of $\bm{z}$, and $E^1_i(\bm{z}) = z_i$, with $c^{1 \to 0}$ being a constant function that always returns $a$.
    Then, if $E^0$ is correct, we have
    \begin{align*}
        Y_i = y^0_i(a) = y^1_i(e^1) \;\; \text{for all $e^1 \in \tilde{\mathcal{E}}^1_i = \{0, 1\}$ and $i \in [n]$.}
    \end{align*}
    Thus, $\mathcal{N}(2) = [n]$ holds if all individuals are treatment eligible.
    In this case, the entire $[n]$ can be used as the focal units and the corresponding focal assignment set is simply given by $\mathcal{C}^\mathcal{S} = \mathcal{Z}$.
\end{example}

Now, let $T(\bm{z}, \bm{Y}_\mathcal{S}(\bm{z}))$ be some predetermined test statistic, where $\bm{Y}_\mathcal{S}(\bm{z}) = (Y_i(\bm{z}))_{i \in \mathcal{S}}$.
The choice of $T$ will be discussed later.
Under $\mathbb{H}_0$, $\bm{Y}_\mathcal{S}(\bm{z})$ is imputable from $\bm{Y}_\mathcal{S} \equiv \bm{Y}_\mathcal{S}(\bm{Z})$ as long as $\bm{z} \in \mathcal{C}^\mathcal{S}$.
Then, the $p$-value for $T(\bm{z}, \bm{Y}_\mathcal{S}(\bm{z}))$ conditional on $\mathcal{C}^\mathcal{S}$ under $\mathbb{H}_0$ is the probability that the realization of the test statistic under the conditional randomization distribution is at least as extreme as its actual value:
\begin{align}\label{eq:pval_true}
    \text{p}(\bm{Z}, \mathcal{C}^\mathcal{S}) 
    \coloneqq \Pr[ T(\bm{z}^*, \bm{Y}_\mathcal{S}(\bm{z}^*)) \ge T(\bm{Z}, \bm{Y}_\mathcal{S}) \mid \bm{z}^* \in \mathcal{C}^\mathcal{S} ],
\end{align}
where the probability is with respect to $\bm{z}^* \sim \mathbb{P}_{\bm{Z} \mid \bm{Z} \in \mathcal{C}^\mathcal{S}}$.
In practice, it is difficult to exactly compute \eqref{eq:pval_true} because $|\mathcal{C}^\mathcal{S}|$ is typically very large.
Thus, we propose to approximate the $p$-value using the Monte Carlo method:

\begin{algorithm}
    \caption{Randomization Test}\label{alg:test}
    \begin{algorithmic}[1]
        \Require $\bm{Z}, \bm{Y}_\mathcal{S}, \mathbb{P}_{\bm{Z} \mid \bm{Z} \in \mathcal{C}^\mathcal{S}}$
        \Ensure the estimated $p$-value: $\hat p_R$
        \State Compute $T(\bm{Z}, \bm{Y}_\mathcal{S})$
        \For {$r = 1$ to $R$}
            \State Draw $\bm{z}^{(r)}$ independently from $\mathbb{P}_{\bm{Z} \mid \bm{Z} \in \mathcal{C}^\mathcal{S}}$
            \State Compute $T(\bm{z}^{(r)}, \bm{Y}_\mathcal{S}(\bm{z}^{(r)}))$ under $\mathbb{H}_0$
        \EndFor
        \State Compute $\hat p_R \coloneqq \displaystyle \frac{1}{R} \sum_{r = 1}^R \bm{1}\{T(\bm{z}^{(r)}, \bm{Y}_\mathcal{S}(\bm{z}^{(r)})) \ge T(\bm{Z}, \bm{Y}_\mathcal{S})\}$
    \end{algorithmic}
\end{algorithm}

The next theorem states the validity of this testing procedure.\footnote{
    Here, we implicitly assume that $|\mathcal{S}| \ge \kappa$;
    otherwise, the test statistic may not be well defined.
}

\begin{theorem} \label{thm:validity}
    \hfil 
   \begin{enumerate}[(i)]
        \item $\Pr[ \text{p}(\bm{Z}, \mathcal{C}^\mathcal{S}) \le \alpha \mid \bm{Z} \in \mathcal{C}^\mathcal{S}] \le \alpha$ for any $\alpha \in (0, 1)$ under $\mathbb{H}_0$.
        \item $|\hat p_R - \text{p}(\bm{Z}, \mathcal{C}^\mathcal{S})| = O_P\left(R^{-1/2}\right)$.
    \end{enumerate}
\end{theorem}

In Theorem \ref{thm:validity}(i), the probability of a type I error is generally not precisely the nominal level $\alpha \in (0, 1)$.
This is a common feature of the randomization approach owing to the discrete nature of $\bm{Z}$.
Theorem \ref{thm:validity}(ii) shows that the stochastic order of the Monte Carlo approximation error is $R^{-1/2}$, where the probability is with respect to $\{\bm{z}^{(r)}\} \sim \mathbb{P}_{\bm{Z} \mid \bm{Z} \in \mathcal{C}^\mathcal{S}}$.
Note that this result is independent of the size of the focal subpopulation.
Because $R$ can be freely chosen by researchers, the $p$-value can be estimated with arbitrary precision.

\paragraph{Proof of Theorem \ref{thm:validity}}

(i) By the definition of $\mathcal{C}^\mathcal{S}$, for any $\bm{z}^* \in \mathcal{C}^\mathcal{S}$, we have $\bm{Y}_\mathcal{S}(\bm{z}^*) = \bm{Y}_\mathcal{S}$ under $\mathbb{H}_0$.
Thus, we can write $\text{p}(\bm{Z}, \mathcal{C}^\mathcal{S}) = \Pr[ T(\bm{z}^*, \bm{Y}_\mathcal{S}) \ge T(\bm{Z}, \bm{Y}_\mathcal{S}) \mid \bm{z}^* \in \mathcal{C}^\mathcal{S} ]$, where the probability is with respect to $\bm{z}^* \sim \mathbb{P}_{\bm{Z} \mid \bm{Z} \in \mathcal{C}^\mathcal{S}}$.
Let $F_{T \mid \mathcal{C}^\mathcal{S}}$ denote the conditional distribution function of $-T(\bm{z}^*, \bm{Y}_\mathcal{S})$ given $\bm{z}^* \in \mathcal{C}^\mathcal{S}$ induced from $\mathbb{P}_{\bm{Z} \mid \bm{Z} \in \mathcal{C}^\mathcal{S}}$.
Then, $\text{p}(\bm{Z}, \mathcal{C}^\mathcal{S}) = F_{T \mid \mathcal{C}^\mathcal{S}}(-T(\bm{Z}, \bm{Y}_\mathcal{S}))$, and, thus,
\begin{align*}
    \Pr[\text{p}(\bm{Z}, \mathcal{C}^\mathcal{S}) \le \alpha \mid \bm{Z} \in \mathcal{C}^\mathcal{S}] 
    = \Pr[F_{T \mid \mathcal{C}^\mathcal{S}}(-T(\bm{Z}, \bm{Y}_\mathcal{S})) \le \alpha \mid \bm{Z} \in \mathcal{C}^\mathcal{S}]
    \le \alpha,
\end{align*}
where the inequality follows from the fact that $-T(\bm{Z}, \bm{Y}_\mathcal{S})$ is distributed as $F_{T \mid \mathcal{C}^\mathcal{S}}$ given $\bm{Z} \in \mathcal{C}^\mathcal{S}$.

\bigskip
\noindent 
(ii) Let $p(\bm{z}^{(r)}) \coloneqq \bm{1}\{T(\bm{z}^{(r)}, \bm{Y}_\mathcal{S}) \ge T(\bm{Z}, \bm{Y}_\mathcal{S})\}$, such that $\hat p_R = R^{-1} \sum_{r = 1}^R p(\bm{z}^{(r)})$.
Because $\bm{z}^{(r)}$'s are identically drawn from $\mathbb{P}_{\bm{Z} \mid \bm{Z} \in \mathcal{C}^\mathcal{S}}$, we have $\mathbb{E}_{\{\bm{z}^{(r)}\} \sim \mathbb{P}_{\bm{Z} \mid \bm{Z} \in \mathcal{C}^\mathcal{S}}} \hat p_R = \mathbb{E}_{\bm{z} \sim \mathbb{P}_{\bm{Z} \mid \bm{Z} \in \mathcal{C}^\mathcal{S}}} p(\bm{z}) = \text{p}(\bm{Z}, \mathcal{C}^\mathcal{S})$.
Furthermore, by the independence of the draws,
\begin{align*}
    \mathbb{E}_{\{\bm{z}^{(r)}\} \sim \mathbb{P}_{\bm{Z} \mid \bm{Z} \in \mathcal{C}^\mathcal{S}}} \left( \hat p_R - \text{p}(\bm{Z}, \mathcal{C}^\mathcal{S}) \right)^2
    = \frac{\mathbb{E}_{\bm{z} \sim \mathbb{P}_{\bm{Z} \mid \bm{Z} \in \mathcal{C}^\mathcal{S}}} ( p(\bm{z}) - \text{p}(\bm{Z}, \mathcal{C}^\mathcal{S}) )^2}{R}
    = \frac{\text{p}(\bm{Z}, \mathcal{C}^\mathcal{S})(1 - \text{p}(\bm{Z}, \mathcal{C}^\mathcal{S}))}{R}.
\end{align*}
Thus, the result follows from Chebyshev's inequality. \qed

\begin{remark}[Sampling from $\mathbb{P}_{\bm{Z} \mid \bm{Z} \in \mathcal{C}^\mathcal{S}}$] \label{remark:sampling}
    Procedure \ref{alg:test} requires repeatedly sampling new $\bm{z}$'s from $\mathbb{P}_{\bm{Z} \mid \bm{Z} \in \mathcal{C}^\mathcal{S}}$.
    In certain special cases, for example, when $E^0_i(\bm{z}) = z_i$ and $\mathbb{P}_{\bm{Z}}$ is given by Bernoulli trials, we can draw directly from $\mathbb{P}_{\bm{Z} \mid \bm{Z} \in \mathcal{C}^\mathcal{S}}$ relatively easily.
    Even when $E^0$ is of a more general form, noting that $\mathbb{P}_{\bm{Z} \mid \bm{Z} \in \mathcal{C}^\mathcal{S}}(\bm{z}) \, \propto \, \bm{1}\{\bm{z} \in \mathcal{C}^\mathcal{S}\} \cdot \mathbb{P}_{\bm{Z}}(\bm{z})$, sampling from $\mathbb{P}_{\bm{Z} \mid \bm{Z} \in \mathcal{C}^\mathcal{S}}$ can be done manually by preliminarily drawing $\bm{z}$ from $\mathbb{P}_{\bm{Z}}$, and if it satisfies $\bm{z} \in \mathcal{C}^\mathcal{S}$, we keep this $\bm{z}$ and move on to the computation of the test statistic; otherwise, we re-draw a new $\bm{z}$ from $\mathbb{P}_{\bm{Z}}$.
    Although this approach is technically simple, it has a drawback in that, if $\mathcal{Z}$ is a huge set and $\mathcal{C}^\mathcal{S}$ is small relative to $\mathcal{Z}$, the probability of observing $\bm{z}$ satisfying $\bm{z} \in \mathcal{C}^\mathcal{S}$ can be extremely small, which makes it computationally very inefficient.
    This computational issue is left for future work.
\end{remark}

\begin{remark}[Choice of $E^1$]
	In practice, there may be a large number of possible candidates for $E^1$.
	As long as $E^1 \supset E^0$, in terms of size control, the selection of $E^1$ can be arbitrary; however, it may significantly affect the power of the test.
    In general, it would be better to employ a coarser $E^1$ to secure the size of the focal subpopulation.
    However, note that, if $E^1$ is too ``similar'' to $E^0$, the test may not exhibit sufficient power.
    For example, when we test for the presence of treatment spillovers using $E^0_i(\bm{z}) = z_i$ and $E_i^1(\bm{z}) = (z_i, z_1)$, this would result in very low power to detect the spillover effect because only those affected by unit $1$ can contribute to the detection. 
    Thus, ideally, we would like to choose $E^1$ such that it can nicely capture the true interference pattern in a way that $E^0$ cannot, while maintaining its coarseness.
    How to find such an ideal $E^1$ is also left as an important open question.
\end{remark}

\subsection{Construction of the focal subpopulation in a social network framework}\label{subsec:socnet}

If the structure of the population is as simple as that in Example \ref{ex:pair}, the construction of the focal subpopulation $\mathcal{S}$ is straightforward.
However, when one deals with a more general network structure, one finds that forming an appropriate $\mathcal{S}$ is a challenging task.
To address this issue, \cite{athey2018exact} proposed several approaches that systematically or randomly choose focal units based on the shape of the interaction network of each unit, independent of the actual treatment assignment.
However, as pointed out by \cite{basse2019randomization} and \cite{puelz2022graph}, constructing the focal subpopulation without utilizing the observed treatment assignment may result in the loss of the power of the test.
In this subsection, we discuss this issue further in an empirically common social network setup.

Suppose that the individuals are connected through social networks.
Let $\bm{A} = (A_{ij})_{i,j \in [n]}$ be the adjacency matrix, where $A_{ij} \in \{ 0, 1 \}$ represents whether $j$ affects $i$ (directed networks are allowed).
We set $A_{ii} = 0$ for all $i$.
For each $i$, the set of interacting peers is denoted as $\mathcal{P}_i \coloneqq \{j \in [n] : A_{ij} = 1\}$ and individual $i$'s neighborhood is denoted as $\overline{\mathcal{P}}_i \coloneqq \{i\}\cup\mathcal{P}_i$.
For simplicity, assume that the exposure mapping of interest depends only on the individual's own and peers' treatments: $E_i^0(\bm{z}) = E_i^0((z_j)_{j \in \overline{\mathcal{P}}_i})$.

\begin{example}\label{ex:socnet}
    Suppose we have $n = 8$ individuals in the population and that they form an undirected social network, as shown in Figure \ref{fig:socnet}.
    In the figure, the treated individuals are grayed and the controls are white.
    We would like to test whether $E^0_i(\bm{z}) = \max_{j \in \overline{\mathcal{P}}_i} z_j$ is correct, which claims that having at least one treated unit in one's own neighborhood is only important.
    As a finer counterpart of this, let $E^1_i(\bm{z}) = (z_i, \max_{j \in \mathcal{P}_i} z_j)$.

    \begin{figure}[ht]
        \centering
        \includegraphics[bb = 0 0 205 54, width = 7cm]{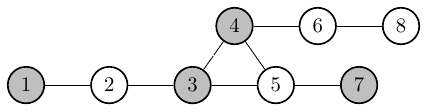}

        \footnotesize
        Note: Gray and white nodes represent the treatment and control units, respectively.
        \caption{Undirected social network $\bm{A}$}
        \label{fig:socnet}
    \end{figure}
    \begin{table}[!ht]
        \begin{center}
            \caption{Potential outcomes schedule under $\mathbb{H}_0$: $E^0$ is correct} \label{table:schedule}
            \small
            \resizebox{\columnwidth}{!}{\begin{tabular}{cccccccccccc}
            \toprule
            ID & $Y_i$ & $Z_i$ & $E_i^0(\bm{Z})$ & $E_i^1(\bm{Z})$ & $y_i^0(0)$ & $y_i^0(1)$ & $y_i^1(0, 0)$ & $y_i^1(1, 0)$ & $y_i^1(0, 1)$ & $y_i^1(1, 1)$ & $\mathcal{N}(3)$ \\
            \midrule
            1 & 4 & 1 & 1 & (1, 0) &  & 4 &  & \ul{4} & 4 & 4 & \checkmark \\
            2 & 3 & 0 & 1 & (0, 1) &  & 3 &  & 3 & \ul{3} & 3 & \checkmark \\
            3 & 7 & 1 & 1 & (1, 1) &  & 7 &  & 7 & 7 & \ul{7} & \checkmark \\
            4 & 8 & 1 & 1 & (1, 1) &  & 8 &  & 8 & 8 & \ul{8} & \checkmark \\
            5 & 2 & 0 & 1 & (0, 1) &  & 2 &  & 2 & \ul{2} & 2 & \checkmark \\
            6 & 3 & 0 & 1 & (0, 1) &  & 3 &  & 3 & \ul{3} & 3 & \checkmark \\
            7 & 5 & 1 & 1 & (1, 0) &  & 5 &  & \ul{5} & 5 & 5 & \checkmark \\
            8 & 1 & 0 & 0 & (0, 0) & 1 &  & \ul{1} &  &  &  & \\
            \bottomrule
            \end{tabular}}
        \end{center}
        \footnotesize
        Note: The underlined $y^1$'s are observed potential outcomes.
    \end{table}
    Suppose that we have observed $(4,3,7,8,2,3,5,1)$ for the outcomes of the units.
    Then, the potential outcomes schedule under $\mathbb{H}_0$ can be summarized as in Table \ref{table:schedule}.
    In the table, the blank cells are those not imputable from the observed outcomes.
    $\mathcal{N}(3)$ comprises the individuals excluding ID 8, with $\tilde{\mathcal{E}}^1_i = \{(1,0), (0,1), (1,1)\}$.
    Then, the ``observed'' $y_i^1(1,0)$'s are $\{4,5\}$.
    Similarly, we obtain $\{3,2,3\}$ as the observed values of $y_i^1(0,1)$ and $\{7,8\}$ as those of $y_i^1(1,1)$.
    When $\mathbb{H}_0$ is true, these three samples should have been drawn from the same distribution, which is exactly the argument to be tested with our approach.
\end{example}

\paragraph{$\epsilon$-net}

\begin{figure}[!ht]
\begin{center}
    \centering
    \includegraphics[bb = 0 0 130 110, width = 5cm]{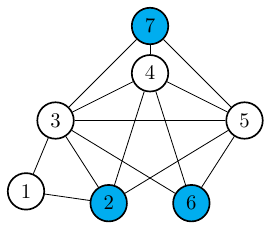}
    \caption{Common-friend graph $\bm{G}$ (Example \ref{ex:socnet})}
    \label{fig:common-friend}
\end{center}
\end{figure}

For the construction of $\mathcal{S}$, a practical approach that generally works for any social network data is to construct $\mathcal{S}$ such that $\overline{\mathcal{P}}_i \cap \overline{\mathcal{P}}_j = \emptyset$ holds for any $i,j \in \mathcal{S}$.
Finding such an $\mathcal{S}$ is a well-known problem in graph theory.
Let $\bm{G} = (\mathcal{N}(\kappa), \bm{E})$ be the ``common-friend'' graph with vertex set $\mathcal{N}(\kappa)$ and edge set $\bm{E} = \{(i,j) \in \mathcal{N}(\kappa) : \overline{\mathcal{P}}_i \cap \overline{\mathcal{P}}_j \ne \emptyset\}$.
Then, an \textit{independent set} of $\bm{G}$, which is a set of vertices such that no two vertices in the set are adjacent, can be a valid candidate for $\mathcal{S}$.
In particular, we would like to find a maximum independent set (MIS) of $\bm{G}$.
Figure \ref{fig:common-friend} shows the common-friend graph for the network data in Example \ref{ex:socnet}.
We have two MIS's, namely, $\{1,6,7\}$ and $\{2,6,7\}$.
When we set $\mathcal{S} = \{2,6,7\}$, the admissible assignment vectors are $\mathcal{C}^\mathcal{S} = \{\bm{z} \in \mathcal{Z}: E_2^0(\bm{z}) = E_6^0(\bm{z}) = E_7^0(\bm{z}) = 1\}$.

The above approach of constructing $\mathcal{S}$ is characterized as the \textit{$\epsilon$-net} of $\mathcal{N}(\kappa)$ with $\epsilon = 3$.
Let $d(i,j)$ be the path distance between $i$ and $j$ on $\bm{A}$.
For a non-negative integer $\epsilon$, define $\mathcal{B}_{\epsilon}(i) \coloneqq \{ j \in \mathcal{N}(\kappa): d(i, j) \le \epsilon \}$ as the subset of $\mathcal{N}(\kappa)$ within $\epsilon$ distance of unit $i$.
An $\epsilon$-net of $\mathcal{N}(\kappa)$ is a set $\mathcal{S}$ of units such that $d(i, j) \ge \epsilon$ for all $i, j \in \mathcal{S}$ and $\mathcal{N}(\kappa) \subseteq \bigcup_{i \in \mathcal{S}} \mathcal{B}_{\epsilon}(i)$.
The $\epsilon$-net was also considered in \cite{athey2018exact} in a similar context, but ours is different from theirs in that we first select $\mathcal{N}(\kappa)$ according to the observed treatments, which potentially results in an improvement in the power of the test.

\bigskip

Alternatively to the 3-net (MIS of $\bm{G}$), one might consider employing the largest 2-net of $\mathcal{N}(\kappa)$ as $\mathcal{S}$.
Note that the largest 2-net is equivalent to the MIS of $\bm{A}$, not of $\bm{G}$.
An advantage of using 2-net is that its size is generally larger than 3-nets.
On the other hand, unlike 3-net, 2-net results in focal units with overlapping peers, which may complicate the testing procedure and lower the power of the test.
In Example \ref{ex:socnet}, the largest 2-net of $\mathcal{N}(3)$ is $\{ 1, 3, 6, 7 \}$, and the corresponding focal assignments are $\mathcal{C}^\mathcal{S} = \{\bm{z} \in \mathcal{Z}: E_1^0(\bm{z}) = E_3^0(\bm{z}) = E_6^0(\bm{z}) = E_7^0(\bm{z}) = 1\}$.

\paragraph{Biclique method}

We can extend the biclique method in \cite{puelz2022graph} to our situation.
To this end, we define the null exposure graph and its biclique in our context.
Let $\mathcal{Z}_0 \subseteq \mathcal{Z}$ denote a predetermined set of treatment assignments such that $\bm{Z} \in \mathcal{Z}_0$.
For example, one may construct $\mathcal{Z}_0$ by drawing from $\mathbb{P}_{\bm{Z}}$ sufficiently many times.
The null exposure graph of $\mathbb{H}_0$ with respect to $\mathcal{Z}_0$ is defined as a bipartite graph $\bm{L} = (\mathcal{N}(\kappa) \cup \mathcal{Z}_0, \bm{E})$, where $\bm{E} = \{ (i, \bm{z}) \in \mathcal{N}(\kappa) \cup \mathcal{Z}_0: c^{1 \to 0}(E_i^1(\bm{z})) = E_i^0(\bm{Z}) \}$.
That is, there exists an edge between $i \in \mathcal{N}(\kappa)$ and $\bm{z} \in \mathcal{Z}_0$ when $Y_i = y_i^0(E_i^0(\bm{Z})) = y_i^1(E_i^1(\bm{z}))$ holds under $\mathbb{H}_0$.
Then, a biclique $\bm{B}_b = (\mathcal{N}_b, \mathcal{Z}_b)$ in $\bm{L}$ is defined as a pair of sets $\mathcal{N}_b \subseteq \mathcal{N}(\kappa)$ and $\mathcal{Z}_b \subseteq \mathcal{Z}_0$ such that $(i, \bm{z}) \in \bm{E}$ holds for all $i \in \mathcal{N}_b$ and $\bm{z} \in \mathcal{Z}_b$.
In general, we can find multiple bicliques for $\bm{L}$ ($b = 1, 2, \ldots$) and it is typically desirable to have a larger biclique.
By construction, if $\mathbb{H}_0$ is true, then we have $Y_i = y_i^0(E_i^0(\bm{Z})) = y_i^1(E_i^1(\bm{z}))$ for all $(i, \bm{z}) \in \bm{B}_b$.
Once a biclique $\bm{B}_b$ is obtained, we can simply set $\mathcal{S} = \mathcal{N}_b$ and $\mathcal{C}^{\mathcal{S}} = \mathcal{Z}_b$.

\subsection{Choice of test statistic}\label{subsec:stat}

There is certain freedom in the choice of test statistics.
We consider the following three types of statistics: Kruskal-Wallis (KW), average cross difference (ACD), and ordinary least squares (OLS).

To define these statistics, we introduce additional notations.
For each $i \in \mathcal{S}$, we order the elements of $\tilde{\mathcal{E}}^1_i$ as $e_1^1, e_2^1, \ldots, e_\kappa^1$ based on some rule.
For example, in the case of Example \ref{ex:socnet}, we can consider an increasing order in terms of the value of $2 z_i + \max_{j \in \mathcal{P}_i} z_j$, leading to $(e_1^1, e_2^1, e_3^1) = ((0, 1), (1, 0), (1, 1))$.
Note that, because $\tilde{\mathcal{E}}^1_i$ may be heterogeneous among individuals, the compositions of the ordered elements $(e_1^1, e_2^1, \ldots, e_\kappa^1)$ are also generally different among these individuals.
When such a heterogeneity is present, what sorting rule is adopted is a factor that affects the power of the test.
For a given treatment assignment $\bm{z}$, we partition the focal subpopulation $\mathcal{S}$ into $\kappa$ groups: $\mathcal{S}_j(\bm{z}) \coloneqq \{i \in \mathcal{S} : E_i^1(\bm{z}) = e_j^1 \}$ for $j \in [\kappa]$.
Now, we have $\kappa$ potential outcomes $(y_i^1(e_1^1), y_i^1(e_2^1), \ldots , y_i^1(e_\kappa^1))$ for each $i$, which should take the same value under $\mathbb{H}_0$.

Noting that our task can be viewed as testing the equivalence of $\kappa$ different treatments, we consider the use of the KW statistic, as in \cite{keele2012strengthening} and \cite{wang2020randomization}.
First, we rank all $(Y_i)_{i \in \mathcal{S}}$ from $1$ to $|\mathcal{S}|$.
Let $v_i$ be the rank of $Y_i$ and $V_j(\bm{z})$ be the summation of the ranks for group $\mathcal{S}_j(\bm{z})$: $V_j(\bm{z}) \coloneqq \sum_{i \in \mathcal{S}_j(\bm{z})} v_i$.
The KW statistic compares the average rank for each group $j$, $V_j(\bm{z}) / |\mathcal{S}_j(\bm{z})|$, with the average rank for the entire $|\mathcal{S}|$, $(|\mathcal{S}| + 1) / 2$:
\begin{align*}
        T(\bm{z}, \bm{Y}_\mathcal{S}) 
        = \frac{12}{|\mathcal{S}|(|\mathcal{S}| + 1)} \sum_{j = 1}^{\kappa} |\mathcal{S}_j(\bm{z})| \left( \frac{V_j(\bm{z})}{|\mathcal{S}_j(\bm{z})|} - \frac{|\mathcal{S}| + 1}{2} \right)^2.
\end{align*}

The ACD statistic is defined simply as the average of the absolute average differences for all combinations of treatment pairs:
\begin{align*}
    T(\bm{z}, \bm{Y}_\mathcal{S}) 
    = \frac{2}{\kappa(\kappa - 1)} \sum_{1 \le j < k \le \kappa} \left| \frac{\sum_{i \in \mathcal{S}_j(\bm{z})} Y_i}{|\mathcal{S}_j(\bm{z})|} - \frac{\sum_{i \in \mathcal{S}_k(\bm{z})} Y_i}{|\mathcal{S}_k(\bm{z})|} \right|.
\end{align*}
    
It is also possible to consider a ``model-based'' test statistic, as in \cite{athey2018exact}.
Suppose we have some $E^0$ and $E^1$, where $E^0$ might be vector-valued, and let $X_i(\bm{z})$ be a vector of variables whose values are determined only through $E^1_i(\bm{z})$ but not through $E^0_i(\bm{z})$.
For example, when $E^0_i(\bm{z}) = z_i$ and $E^1_i(\bm{z}) = (z_i, \max_{j \in \mathcal{P}_i} z_j)$, one may  use $X_i(\bm{z}) = \max_{j \in \mathcal{P}_i} z_j$.
Then, by fitting the following regression model to the data in $\mathcal{S}$,
\begin{align}\label{eq:OLS}
    Y_i = \beta_0 + E^0_i(\bm{z})^\top \beta_1 + X_i(\bm{z})^\top \beta_2 + \text{error}_i,
\end{align}
we can use the $F$-statistic for the significance of $\hat \beta_2$ as the test statistic $T(\bm{z}, \bm{Y}_\mathcal{S})$, where $\hat \beta_2$ denotes the OLS estimate. 

For another example, when we have $E_i^0(\bm{z}) = \max_{j \in \overline{\mathcal{P}}_i} z_j$ and $E^1_i(\bm{z}) = (z_i, \max_{j \in \mathcal{P}_i} z_j)$, as in Example \ref{ex:socnet}, we may consider using $X_i(\bm{z}) = (z_i, \max_{j \in \mathcal{P}_i} z_j)$.
Note that, when one adopts this model-based approach, the presumed model does not have to perfectly reflect the true interference structure.
However, if they are significantly different, it will lead to a substantial loss of the power, as numerically demonstrated in Section \ref{sec:MC}.

\begin{remark}[Multiplicity of test statistics]
    As shown above, we generally have multiple statistics for testing $\mathbb{H}_0$.
    Furthermore, by considering different values of $\kappa$ and $E^1$, we can generate a large number of additional test statistics.
    One simple way to utilize the information in all $s$ different statistics altogether is to combine them into a single test statistic,  $T^\text{comb} = g(T^1, T^2, \ldots, T^s)$, as suggested in \cite{imbens2015causal}.
    Another approach is to apply, for example, Simes' correction for multiple testing: letting the ordered $p$-values be denoted by $\hat p_R^{(1)} \le \cdots \le \hat p_R^{(s)}$, reject $\mathbb{H}_0$ if $\hat p_R^{(i)} \le i\alpha /s$ for some $i \in \{ 1, \dots, s \}$.
    See \cite{simes1986improved} and Subsection 9.2.2 of \cite{lehmann2022testing} for more details.
\end{remark}

\subsection{Imperfect compliance}\label{subsec:imperfect}

Thus far, we have assumed that all individuals comply with their initial treatment assignments.
However, in certain realistic situations, they are allowed to self-select their own treatment status.
Now, we write $\bm{D} = (D_1, \ldots, D_n)$ as the $n$-dimensional vector of the actual treatment take-ups.
When noncompliance is allowed ($\bm{D} \neq \bm{Z}$), the probability distribution of $\bm{D}$ is generally unknown. 
Thus, in this case, we cannot perform the test in Procedure \ref{alg:test} based on the actual treatments because it is infeasible to resample independent copies $\bm{d}^{(r)}$'s of $\bm{D}$ from a known distribution.

One empirically tractable approach to this problem is to resort to an intention-to-treat (ITT) analysis.
That is, we consider formulating the exposure mapping as a function not of $\bm{d}$ but of the initial assignment $\bm{z}$.\footnote{
    This type of exposure mapping was considered in \cite{hoshino2023causal} and is termed as \textit{instrumental exposure mapping}.
}
For example, suppose we have the following treatment choice model:
\begin{align*}
    D_i = \bm{1} \left\{ \gamma_{0i} + \gamma_{1i} Z_i + \gamma_{2i} \sum_{j \in \mathcal{P}_i} Z_j > 0 \right\}
\end{align*}
and there are no treatment spillovers in the outcome model.
In this case, the exposure mapping of interest would be $E^0_i(\bm{z}) = (z_i, \sum_{j \in \mathcal{P}_i} z_j)$.
Then, if we can find an appropriate $E^1$ that is finer than $E^0$, in exactly the same way as in Procedure \ref{alg:test}, we can test the validity of this model specification.

\section{Numerical Simulations}\label{sec:MC}

\subsection{Perfect compliance}

In this section, we assess the small sample performance of our randomization test through Monte Carlo simulations.
First, we consider the case of perfect compliance.
The undirected network is created from a simple Erd\"os--R\'enyi model with a probability of $p = 3/n$, where we set $n = 200$.
We consider the following two data generating processes (DGPs) for the outcome variables:
\begin{align*}
    \textbf{DGP 1:} \; \; Y_i = D_i + \tau \sum_{j \in \mathcal{P}_i} D_j + \xi_i,
    \qquad 
    \textbf{DGP 2:} \; \; Y_i = D_i + \tau \cdot g \left( \sum_{j \in \mathcal{P}_i} D_j \right) + \xi_i, 
\end{align*}
where $\xi_i \sim N(0, 1)$, $\tau \in [0, 2]$, and $g(a) = \bm{1}\{ a \le 2 \} \cdot a + \bm{1}\{ a \ge 3 \} \cdot a^{-1}$.
For the treatment assignment mechanism, we employ a complete randomization, where randomly selected $n/2$ units receive $Z = 1$.
Because perfect compliance is assumed here, $D_i = Z_i$ holds for all $i$.
We set $E_i^0(\bm{z}) = z_i$, and, hence, $E^0$ is correct when $\tau = 0$.
For the choice of $E^1$, the following two exposure mappings are used:
\begin{align*}
	\textbf{Exposure 1:} \;\; E_i^1(\bm{z}) = \left( z_i, \max_{j \in \mathcal{P}_i} z_j \right),
	\qquad 
	\textbf{Exposure 2:} \;\; E_i^1(\bm{z}) = \left( z_i, \sum_{j \in \mathcal{P}_i} z_j \right).
\end{align*}
Note that Exposure 1 is coarser than Exposure 2 and that only Exposure 2 is correct when $\tau > 0$.
For Exposure 1, it is natural to set $\kappa = 2$ such that $\tilde{\mathcal{E}}_i = \{(Z_i, 0), (Z_i, 1)\}$, and $\mathcal{N}(2) = \{ i \in [n]: |\mathcal{P}_i| > 0 \}$.
For Exposure 2, we set $\kappa = 4$ such that $\tilde{\mathcal{E}}_i = \{(Z_i, 0), \ldots, (Z_i, 3)\}$ and $\mathcal{N}(4) = \{ i \in [n]: |\mathcal{P}_i| = 3 \}$.

To construct the focal subpopulation $\mathcal{S}$, we consider the following four approaches:
(i) 3-net (MIS of $\bm{G}$),
(ii) 2-net (MIS of $\bm{A}$),
(iii) random selection of $|\mathcal{N}(\kappa)| / 2$ focal units, and
(iv) biclique.
Here, note that finding the largest independent set and finding the largest biclique are both NP-hard problems, and, thus, we approximate their solutions using a greedy vertex coloring algorithm and the binary inclusion-maximal biclustering method, respectively.\footnote{
    Specifically, in the Monte Carlo simulations and the empirical illustrations below, we use the functions {\ttfamily greedy\_vertex\_coloring()} and {\ttfamily BCBimax()} in the R packages {\ttfamily igraph} and {\ttfamily biclust}, respectively.
    }
For (i)--(iii), we set $R = 2,000$.
For (iv), we draw treatment assignments 8 million times from $\mathbb{P}_{\bm{Z}}$ to create $\mathcal{Z}_0$.

For each setup, we perform our randomization test using the KW, ACD, and OLS statistics under the nominal significance level of 5\%.
The OLS statistic is obtained as the $F$-statistic for the OLS estimate of $\beta_2$ in \eqref{eq:OLS}, where we set $X_i(\bm{z}) = \max_{j \in \mathcal{P}_i} z_j$ for Exposure 1 and $X_i(\bm{z}) = \sum_{j \in \mathcal{P}_i} z_j$ for Exposure 2.
In addition to these three tests, we also report the results from the Simes-corrected $p$-value based on them.
The following results are based on 500 Monte Carlo replications.

\begin{figure}[!t]
    \centering
    \begin{subfigure}[h]{\textwidth}
        \includegraphics[bb = 0 0 1820 796, width=17cm]{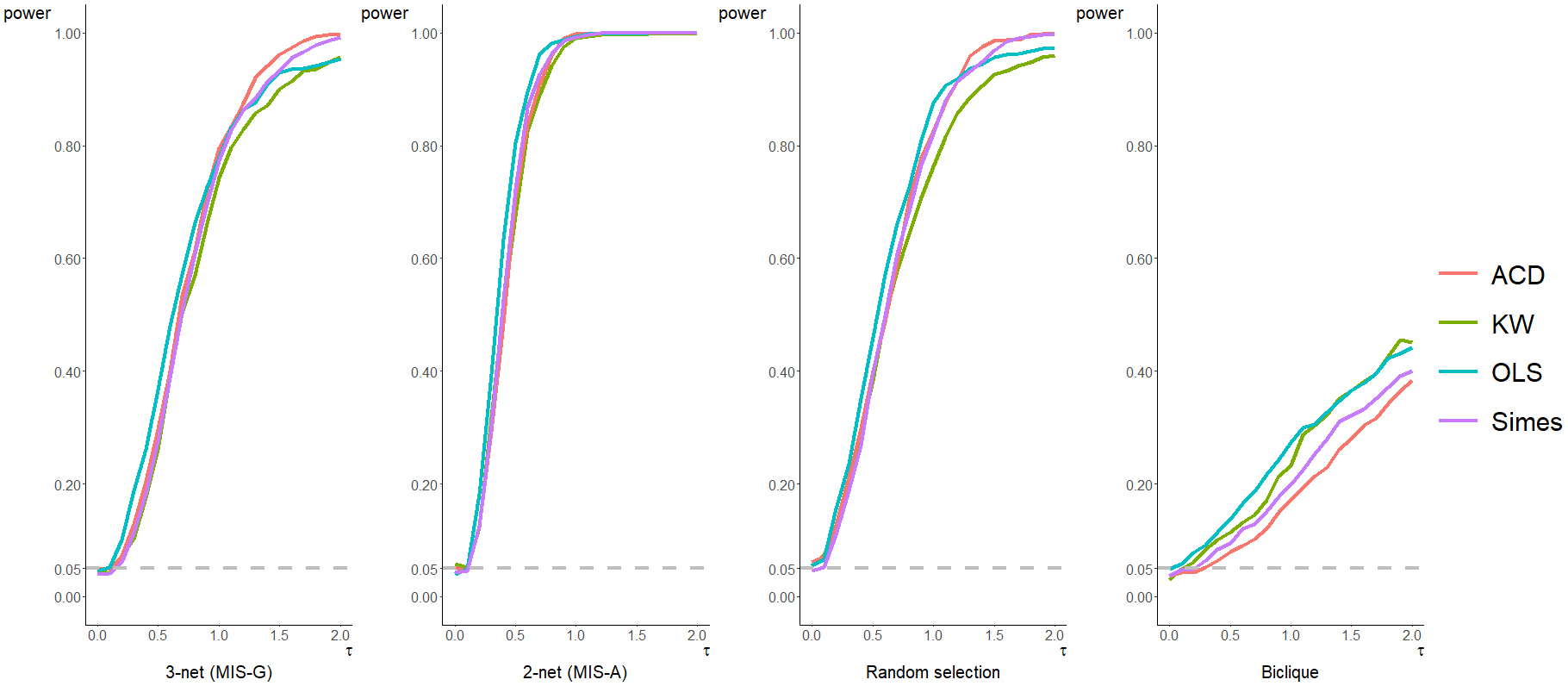}
        \caption{Exposure 1}
    \end{subfigure}
    \begin{subfigure}[h]{\textwidth}
        \includegraphics[bb = 0 0 1820 796, width=17cm]{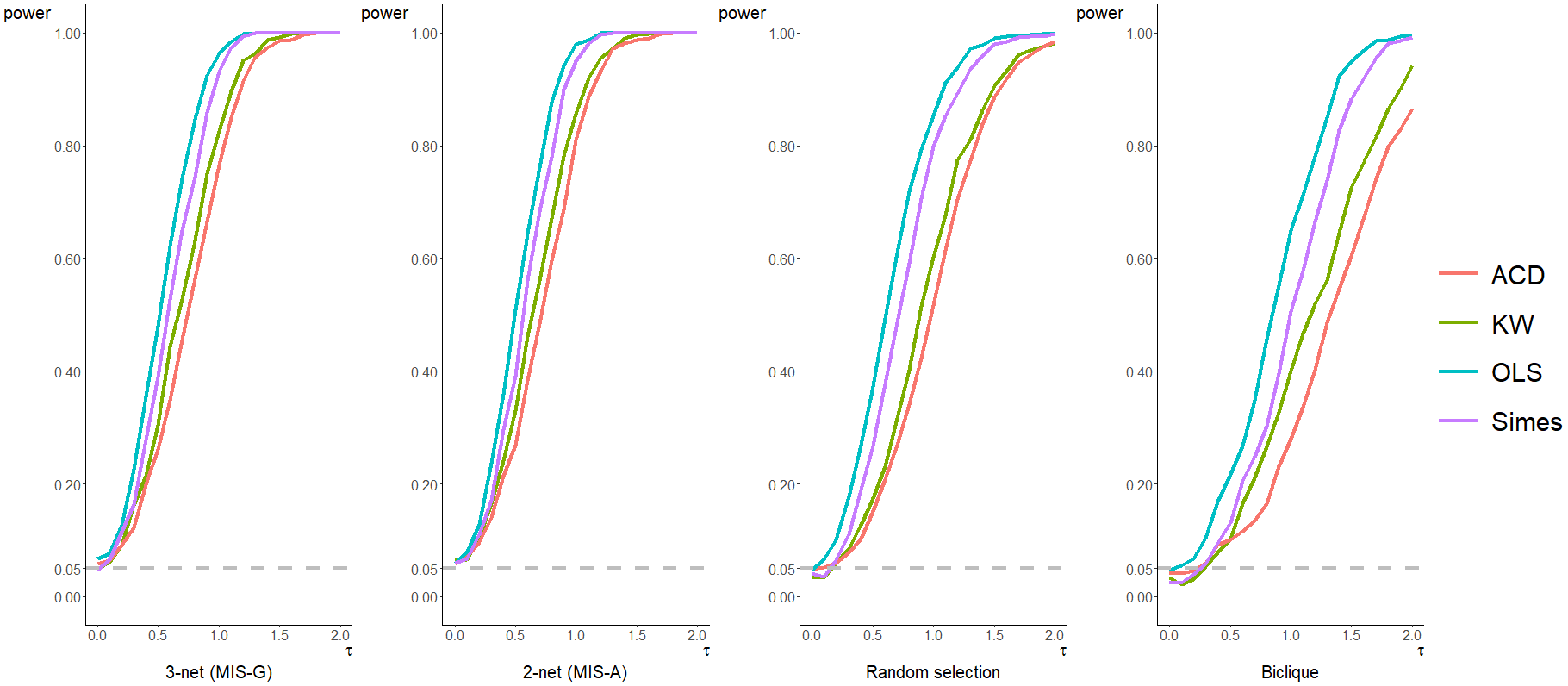}
        \caption{Exposure 2}
    \end{subfigure}
    \caption{Simulation results: DGP 1 under perfect compliance}
    \label{fig:MC1}
\end{figure}

\begin{figure}[!t]
	\centering
	\begin{subfigure}[h]{\textwidth}
		\includegraphics[bb = 0 0 1820 796, width=17cm]{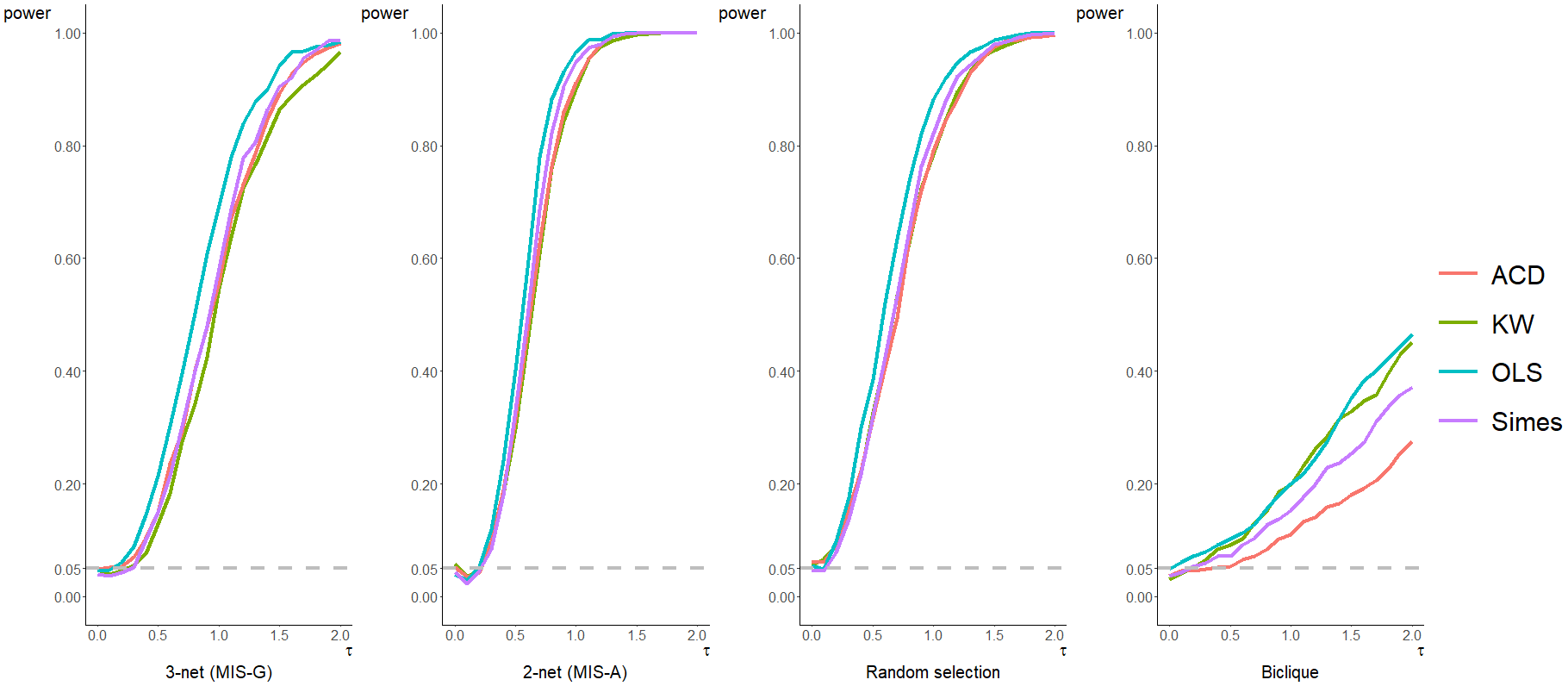}
		\caption{Exposure 1}
	\end{subfigure}
	\begin{subfigure}[h]{\textwidth}
		\includegraphics[bb = 0 0 1820 796, width=17cm]{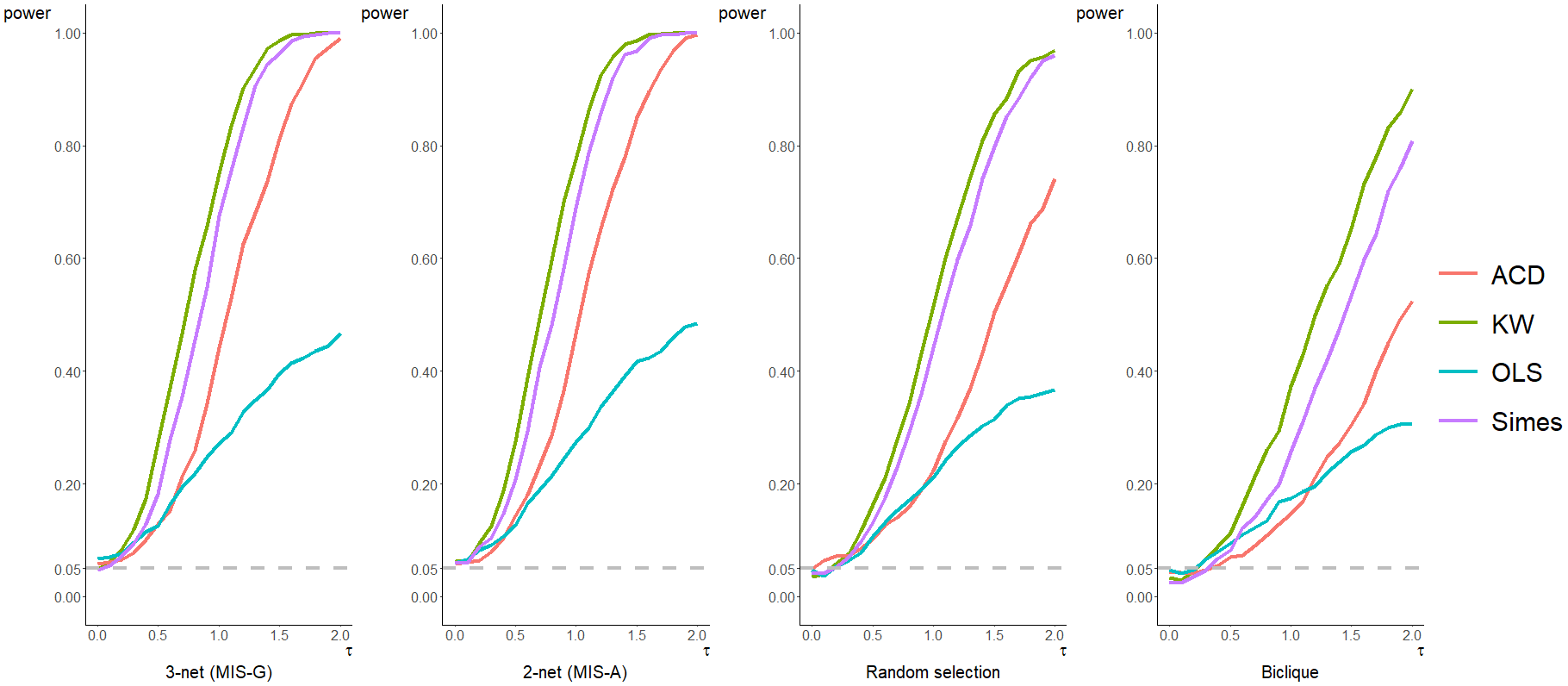}
		\caption{Exposure 2}
	\end{subfigure}
	\caption{Simulation results: DGP 2 under perfect compliance}
	\label{fig:MC2}
\end{figure}

Figures \ref{fig:MC1} and \ref{fig:MC2} show the rejection frequency of each method for different $\tau$ values in DGPs 1 and 2, respectively.
In each figure, panels (a) and (b) present the simulation results for Exposures 1 and 2, respectively.
When $\tau = 0$, for all methods and test statistics, the rejection frequencies are sufficiently close to the nominal level in both DGPs, which is consistent with our theory.
Particularly in DGP 1, the power of these tests quickly increases as $\tau$ increases, suggesting the consistency of our testing procedure.
However, in DGP 2, we find that the power of the OLS statistic based on Exposure 2 is significantly reduced.
This may be due to ``model misspecification'' in the OLS regression caused by the mishandling of the nonlinearity of the $g$ function in this DGP.
Note that the OLS model with Exposure 1 is also a misspecified model; however, the magnitude of the misspecification is mild relative to that of Exposure 2.
Even when Exposure 2 is used in DGP 2, the KW statistic remains sufficiently powerful.

Comparing the four methods for constructing the focal subpopulation, we find that the two MIS-based methods perform better and the biclique the least.
The random selection approach is in-between.
However, a caution should be needed when interpreting this result.
That is, a large part of the difference in the performance of these methods is essentially due to the difference in the sizes of $\mathcal{S}$ and $\mathcal{C}^\mathcal{S}$.
Finding a reasonably large biclique becomes more difficult when the null exposure graph is sparser, as in this simulation setting (see Subsection 6.2 of \citet{puelz2022graph} for a related discussion).
In addition, even with the above-mentioned simplified algorithm, finding a large biclique is still computationally very demanding; for example, even though we have employed a fairly large $\mathcal{Z}_0$, the resulting size of $\mathcal{S}$ was, on average, less than 20 or so after a long computation time.
In a different setup where the biclique method can easily identify relatively large $\mathcal{S}$ and $\mathcal{C}^\mathcal{S}$, its performance would be substantially improved.

For both DGPs, except for the biclique method, Exposure 1 tends to provide more powerful tests than those of Exposure 2, even though Exposure 1 is incorrect when $\tau > 0$.
This result is possibly due to the fact that a coarser Exposure 1 generally induces a larger $\mathcal{S}$ than that of Exposure 2, while retaining a strong correlation with the true interference structure.
For example, when the 2-net is used, the sizes of $\mathcal{S}$ generated from Exposures 1 and 2 are approximately 90 and 30, respectively.

\subsection{Imperfect compliance}

Next, we turn to the experiments for the case of imperfect compliance.
In particular, we consider a one-sided compliance situation; that is, only when $Z_i = 1$ can $i$ choose $D_i = 1$.
The initial treatment assignment $\bm{Z}$ is generated from a complete randomization such that $n / 2$ units are eligible to take the treatment.
For treatment-eligible units, the compliance status follows $\mathrm{Bernoulli}(0.8)$ uniformly (i.e., no interference within the treatment choices).
All other parts of the simulation design are the same as those in the perfect compliance case.
Note that $E_i^0(\bm{z}) = z_i$ is correct when $\tau = 0$. 

\begin{figure}[!t]
	\centering
	\begin{subfigure}[h]{\textwidth}
		\includegraphics[bb = 0 0 1820 796, width=17cm]{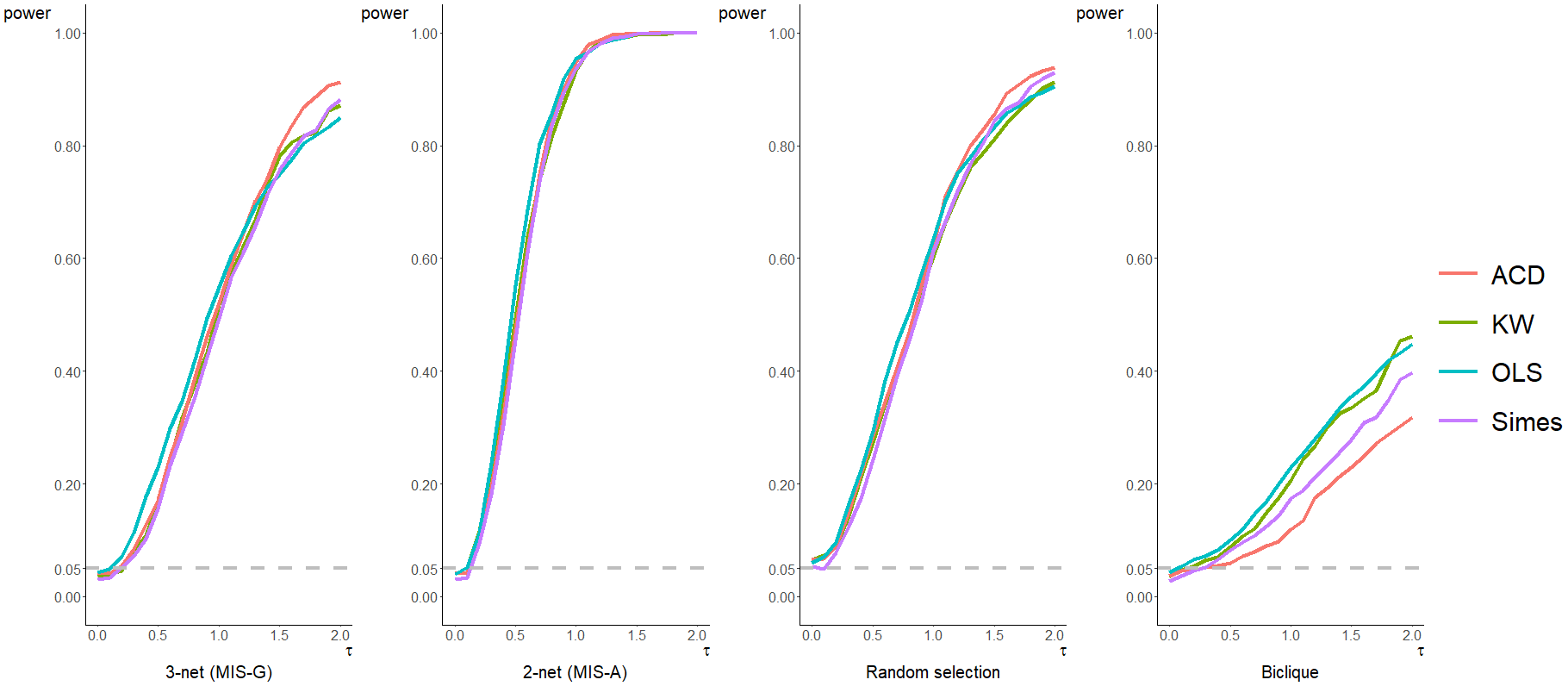}
		\caption{Exposure 1}
	\end{subfigure}
	\begin{subfigure}[h]{\textwidth}
		\includegraphics[bb = 0 0 1820 796, width=17cm]{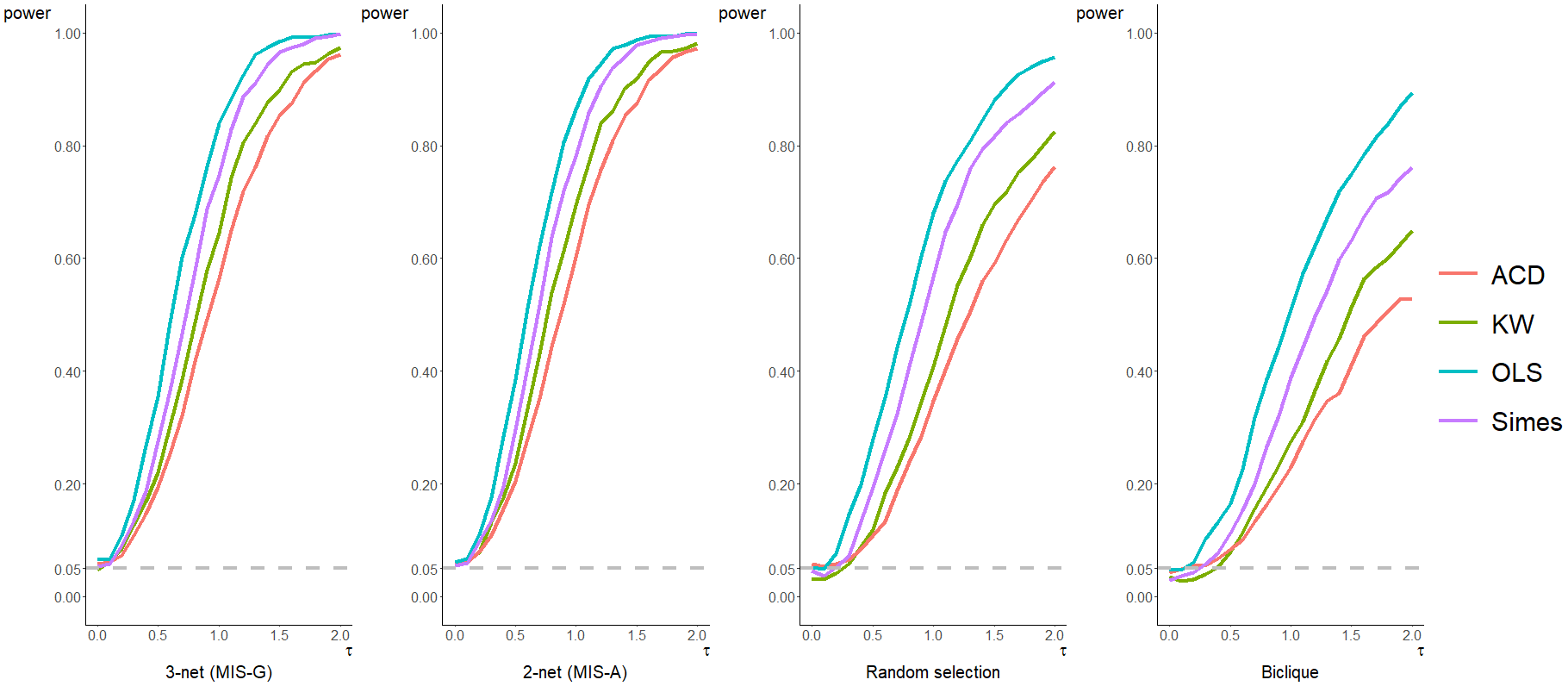}
		\caption{Exposure 2}
	\end{subfigure}
	\caption{Simulation results: DGP 1 under imperfect compliance}
	\label{fig:MC3}
\end{figure}

\begin{figure}[!t]
	\centering
	\begin{subfigure}[h]{\textwidth}
		\includegraphics[bb = 0 0 1820 796, width=17cm]{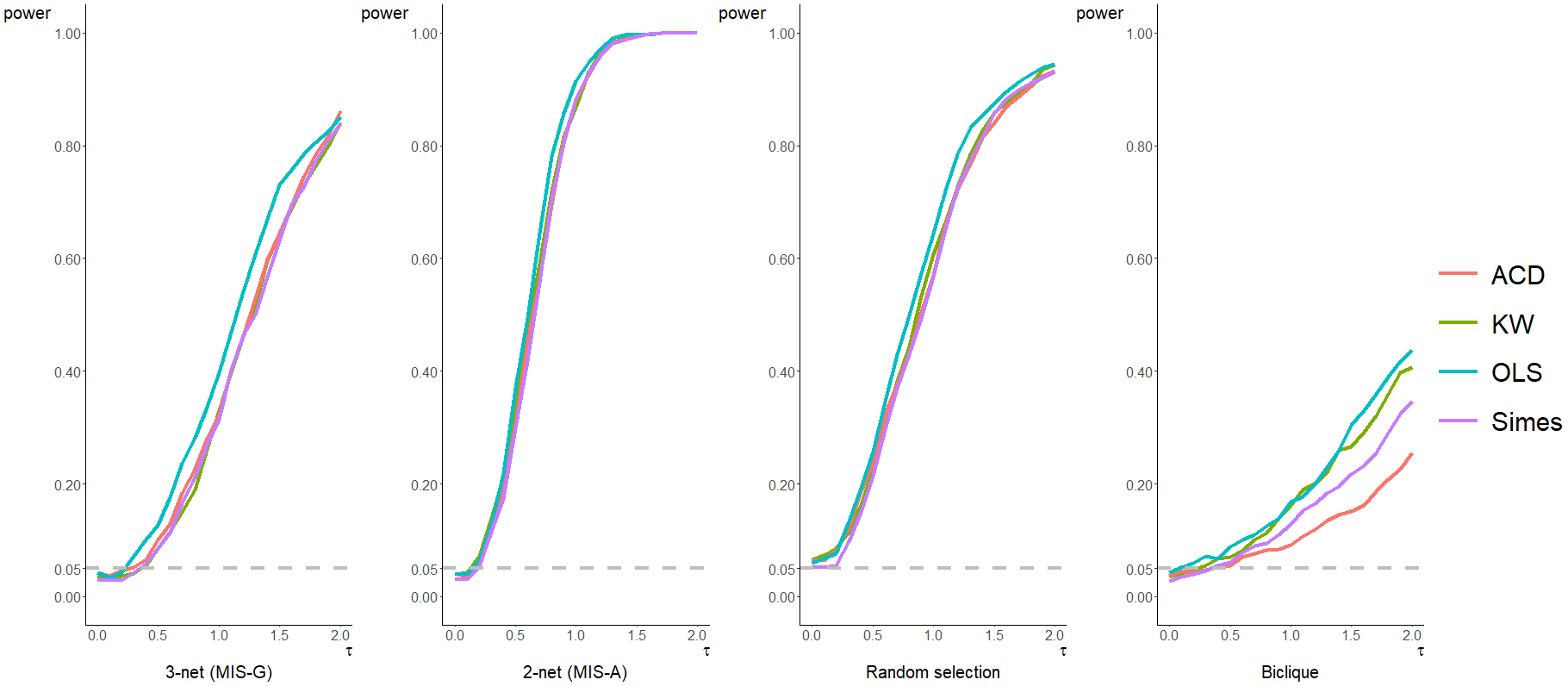}
		\caption{Exposure 1}
	\end{subfigure}
	\begin{subfigure}[h]{\textwidth}
		\includegraphics[bb = 0 0 1820 796, width=17cm]{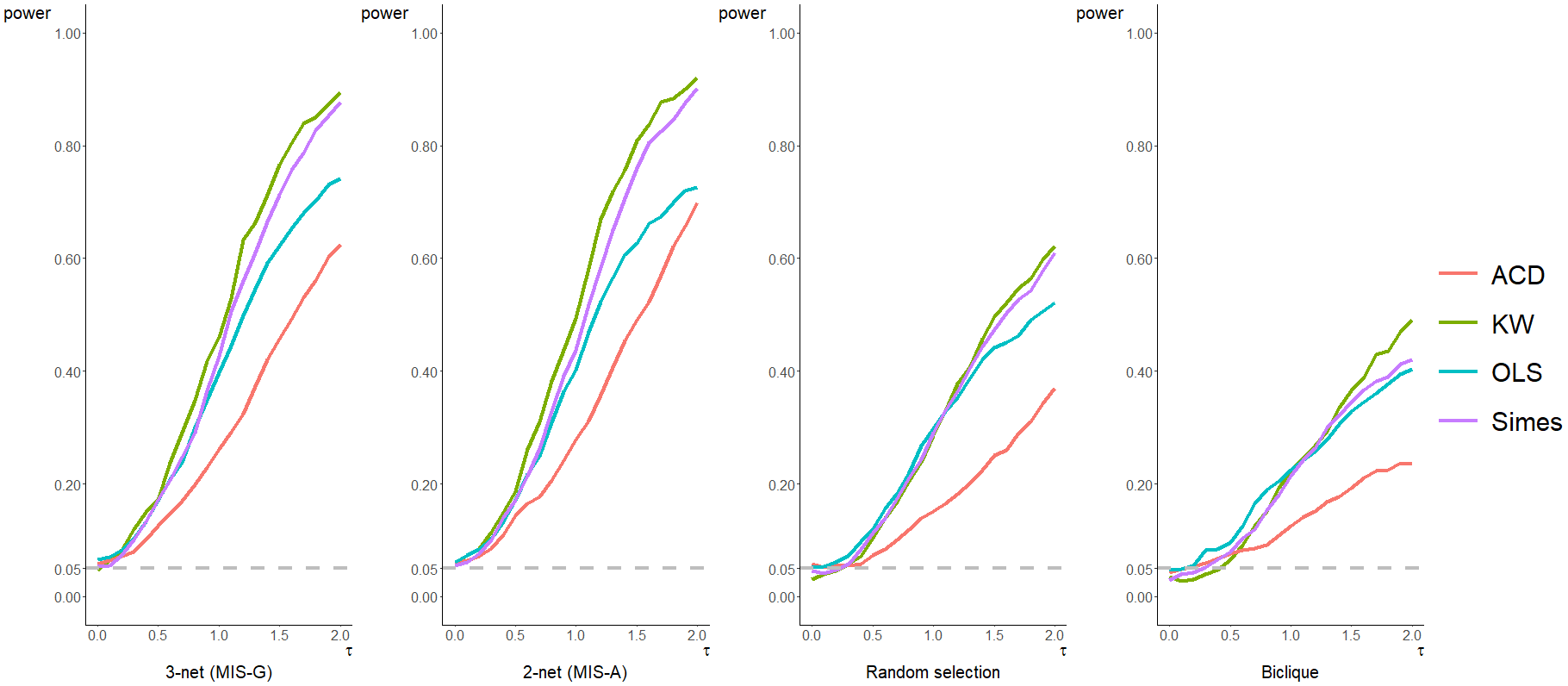}
		\caption{Exposure 2}
	\end{subfigure}
	\caption{Simulation results: DGP 2 under imperfect compliance}
	\label{fig:MC4}
\end{figure}

The simulation results are shown in Figures \ref{fig:MC3} and \ref{fig:MC4}.
Overall, the same comments as in the previous experiment apply to this experiment as well.
Our randomization test works satisfactorily in terms of both size control and power property, although the power of the test seems slightly worse than that in the perfect compliance case.
This is reasonable considering that, unlike in the previous case, neither Exposure 1 nor Exposure 2 is correct when $\tau > 0$.
For the construction of the focal subpopulation, it seems desirable to use a MIS-based method.
An interesting finding is that, when Exposure 2 is employed in DGP 2, it is the ACD statistic, not the OLS statistic, that loses its power significantly.

\section{Empirical Illustrations} \label{sec:empiric}

As empirical illustrations, we apply our randomization test to the existing datasets from two well-known social network experiments in the literature.
The first one is the data on farmers' insurance adoption in \cite{cai2015social}, and the other is the data on anti-conflict intervention school programs in \cite{paluck2016changing}.

For both datasets, we investigate the same type of null hypotheses.
Here, let
\begin{align*}
    E_i^{\mathtt{a}}(\bm{z}) = z_i, 
    \qquad
    E_i^{\mathtt{b}}(\bm{z}) = \max_{j \in \overline{\mathcal{P}}_i} z_j, 
    \qquad
    E_i^{\mathtt{c}}(\bm{z}) = \left( z_i, \max_{j \in \mathcal{P}_i} z_j \right), 
    \qquad
    E_i^{\mathtt{d}}(\bm{z}) = \left( z_i, \sum_{j \in \mathcal{P}_i} z_j \right),
\end{align*}
such that $E^{\mathtt{a}} \subset E^{\mathtt{c}}$, $E^{\mathtt{b}} \subset E^{\mathtt{c}}$, and  $E^{\mathtt{c}} \subset E^{\mathtt{d}}$ hold. 
Table \ref{tab:empirical} summarizes the null hypotheses of interest, the construction of $\mathcal{N}(\kappa)$, and $\tilde{\mathcal{E}}_i^1$ considered in both empirical applications.

For example, the null hypothesis $\mathbb{H}^{\mathtt{a}}$ claims that $E^{\mathtt{a}}$ (no treatment spillovers) is a correct exposure.
To test this null hypothesis, we employ $E^{\mathtt{c}}$ as $E^1$, which contains the information about one's own treatment and whether he/she has at least one treated peer.
This choice of $E^1$ leads to $\tilde{\mathcal{E}}_i^{1} = \{ (Z_i, 0), (Z_i, 1) \}$ with $\kappa = 2$ and to $\mathcal{N}(2) = \{ i \in [n]: |\mathcal{P}_i| > 0 \}$.
The other null hypotheses, $\mathbb{H}^{\mathtt{b}}$ and $\mathbb{H}^{\mathtt{c}}$, can be interpreted similarly.
When testing $\mathbb{H}^{\mathtt{c}}$, we choose a value for $\kappa$ such that the resulting $|\mathcal{S}|$ is maximized.
In the following, given the simulation results in Section \ref{sec:MC}, we report only the results obtained using the two MIS methods.

\begin{table}[!ht]
	\caption{The null hypotheses and settings} \label{tab:empirical}
	\small
	\begin{center}
		\begin{tabular}{ccclcl}
			\toprule
			\multicolumn{1}{c}{$\mathbb{H}_0$} & \multicolumn{1}{c}{$E^0$} & \multicolumn{1}{c}{$E^1$} & \multicolumn{1}{c}{$\tilde{\mathcal{E}}_i^1$} & \multicolumn{1}{c}{$\kappa$} & \multicolumn{1}{c}{$\mathcal{N}(\kappa)$} \\
			\midrule 
			$\mathbb{H}^{\mathtt{a}}$ & $E^{\mathtt{a}}$ & $E^{\mathtt{c}}$ & $\{ (Z_i, 0), (Z_i, 1) \} $    & $\kappa = 2$ & $\{ i \in [n]: |\mathcal{P}_i| > 0 \}$  \\
			$\mathbb{H}^{\mathtt{b}}$ & $E^{\mathtt{b}}$ & $E^{\mathtt{c}}$ & $\{ (0, 1), (1, 0), (1, 1) \}$ & $\kappa = 3$ & $\{ i \in [n]: |\mathcal{P}_i| > 0, E_i^{\mathtt{b}}(\bm{Z}) = 1 \}$  \\
			$\mathbb{H}^{\mathtt{c}}$ & $E^{\mathtt{c}}$ & $E^{\mathtt{d}}$ & $\{ (Z_i, 1), \dots, (Z_i, \kappa) \}$    & $\kappa \ge 2$ & $\{ i \in [n]: |\mathcal{P}_i| = \kappa, E_i^{\mathtt{c}}(\bm{Z}) = (Z_i, 1)\}$ \\
			\bottomrule
		\end{tabular}
	\end{center}
\end{table}

\subsection{Social networks and farmers' insurance decisions} \label{subsec:cai}

\cite{cai2015social} conducted a field experiment to estimate the effect of providing intensive information sessions about the weather insurance on farmers' insurance take-up decisions.
The authors demonstrated the presence of significant treatment spillovers among the farmers by employing a series of regression models,
where the main explanatory variable was the fraction (or number) of friends who were assigned to the intensive information sessions in advance of the focal farmers.

In the experiment, four types of sessions were conducted: first-round simple sessions, first-round intensive sessions, second-round simple sessions, and second-round intensive sessions.
In each round, the simple sessions briefly described the insurance contract, whereas the intensive sessions explained the insurance contract and the expected benefits of the insurance in detail.
To randomly assign the farmers to each session, the authors performed a stratified randomization with four strata constructed in each village according to the household size and rice production acreage.
All rice-producing households were invited to participate in one of the four sessions, and almost 90\% of them attended.
Thus, because the probability of non-attendance was low, we ignore the treatment non-compliance for simplicity.

In this analysis, the outcome variable of interest is $Y_i \in \{ 0, 1 \}$, which indicates whether farmer $i$ decides to buy the weather insurance after attending the session.
Let $\text{int}_i \in \{ 0, 1 \}$ denote whether $i$ is assigned to an intensive session, and let $\text{sec}_i \in \{ 0, 1 \}$ denote whether $i$ is assigned to the second-round session.
Because the treatment spillovers matter only for the participants in the second-round session, as they can receive information from the first-round participants, we create the focal units using only the farmers assigned to the second round.
In addition, to generate sufficient variation in the  $E^1$ values for focal units in testing null hypotheses $\mathbb{H}^{\mathtt{a}}$ and $\mathbb{H}^{\mathtt{b}}$, we restrict the focal subpopulation to be composed of farmers who have fewer than 10 friends.
For the definition of the treatment variable, for a focal unit $i$, we set $z_i = 1$ if $\text{int}_i = 1$.
For a nonfocal unit $j$, we set $z_j = 1$ if both $\text{int}_j = 1$ and $\text{sec}_j = 0$ are true.
When performing our randomization tests, we randomize both $\text{int}_i$ and $\text{sec}_i$ following the protocol of the original experiment.

Table \ref{tab:cai} summarizes the results of our randomization tests.
First, we can see that all $p$-values for testing $\mathbb{H}^{\mathtt{a}}$ are smaller than 5\%, which indicates the presence of information spillovers among the farmers.
For the null hypothesis $\mathbb{H}^{\mathtt{b}}$, two out of six test statistics reject this hypothesis at the 5\% significance level.
Lastly, for $\mathbb{H}^{\mathtt{c}}$, we cannot reject this under any reasonable significance level with any of the test statistics considered.\footnote{
	For $\mathbb{H}^{\mathtt{c}}$, the two MIS methods produce exactly the same focal subpopulation.
	This occurred because of the sparsity of the adjacency matrix restricted on $\mathcal{N}(\kappa)$, which is a phenomenon specific to this particular setup.
}
In summary, these results suggest the existence of spillover effects in farmers' insurance purchasing decisions and that having at least one friend assigned to the intensive session might be an important factor that accounts for the spillovers rather than the number of such friends.\footnote{
    Given this result, one might want to adopt $E^\mathtt{c}$ as the final model and reanalyze the data. However, note that doing so raises another issue of ``inference after model selection'', which is beyond the scope of this paper.
}

\begin{table}
	\caption{Empirical results: farmers' insurance decisions}
	\label{tab:cai}
	\small
	\begin{center}
		\begin{tabular}[t]{rrrrrrrrrr}
			\toprule
			\multicolumn{2}{c}{ } & \multicolumn{3}{c}{$p$-values} & \multicolumn{3}{c}{Simes' correction} & \multicolumn{2}{c}{ } \\
			\cmidrule(l{3pt}r{3pt}){3-5} \cmidrule(l{3pt}r{3pt}){6-8}
			\multicolumn{1}{r}{$\mathcal{S}$} & \multicolumn{1}{c}{$\kappa$} & \multicolumn{1}{c}{KW} & \multicolumn{1}{c}{ACD} & \multicolumn{1}{c}{OLS} & \multicolumn{1}{c}{10\%} & \multicolumn{1}{c}{5\%} & \multicolumn{1}{c}{1\%} & \multicolumn{1}{c}{$|\mathcal{S}|$} & \multicolumn{1}{c}{$R$} \\
			\midrule
			\addlinespace[1em]
			\multicolumn{10}{l}{\underline{Testing for $\mathbb{H}^{\mathtt{a}}$}}\\
			\hspace{2.5cm} 3-net & 2 & 0.006 & 0.014 & 0.006 & \checkmark & \checkmark &  & 507 & 100,000 \\
			\hspace{2.5cm} 2-net & 2 & 0.008 & 0.021 & 0.009 & \checkmark & \checkmark &  & 669 & 100,000 \\
			\addlinespace[1em]
			\multicolumn{10}{l}{\underline{Testing for $\mathbb{H}^{\mathtt{b}}$}}\\
			\hspace{2.5cm} 3-net & 3 & 0.064 & 0.182 & 0.064 & \checkmark &  &  & 485 & 100,000 \\
			\hspace{2.5cm} 2-net & 3 & 0.028 & 0.148 & 0.028 & \checkmark & \checkmark &  & 591 & 100,000 \\
			\addlinespace[1em]
			\multicolumn{10}{l}{\underline{Testing for $\mathbb{H}^{\mathtt{c}}$}}\\
			\hspace{2.5cm} 3-net & 6 & 0.953 & 0.826 & 0.755 &  &  &  & 157 & 100,000 \\
			\hspace{2.5cm} 2-net & 6 & 0.953 & 0.826 & 0.755 &  &  &  & 157 & 100,000 \\
			\bottomrule
		\end{tabular}
	\end{center}
\end{table}

\subsection{Spillover effects of the anti-conflict intervention programs} \label{subsec:paluck}

In the second empirical case study, we apply the proposed test to the data from \citet{paluck2016changing}, who investigated the impact of anti-conflict intervention programs on adolescents' norms and attitudes through a large-scale experiment in 56 American middle schools.
Half of these schools were randomly selected to host the programs.
Within each selected school, a group of students (called {\it seed-eligible students}) were non-randomly selected, and half of these students (called {\it seed students}) were chosen through a stratified randomization and invited to join the program.
The seed-eligible students' strata were determined by their individual characteristics, such as gender, grade, and friendship network variables.
The students' friendship networks were measured by simply asking them to nominate up to 10 friends in their school.
Participation in the program was not mandatory for seed students, and this empirical scenario corresponds to the case of imperfect compliance.
For more details on the experimental design, see the Supplementary Appendix of \citet{paluck2016changing}.

The purpose of this experiment is to examine how the seed students who participated in the intervention program could influence other students through their social networks to improve the climate of the school.
In each intervention meeting, the seed students were encouraged to identify common conflict behaviors in their schools and discuss behavioral strategies to mitigate the conflicts.
As an important role of the seed students, they were allowed to hand out a program wristband as a reward to students for their engagement in friendly or conflict-mitigating behaviors.
Let $Y_i \in \{ 0, 1 \}$ be an indicator of whether student $i$ wears a program wristband, which is the outcome variable of interest in this empirical analysis. 
Let $Z_i \in \{ 0, 1 \}$ indicate the treatment eligibility (i.e., whether $i$ is a seed student).

Table \ref{tab:paluck} summarizes the results of randomization tests to examine the spillover effects of being selected as a seed student on wristband wearing, which can be viewed as the ITT-type analysis discussed in Subsection \ref{subsec:imperfect}.
We can see that all $p$-values for $\mathbb{H}^{\mathtt{a}}$ are sufficiently small to reject them at the 5\% significance level.
The rejection of the hypothesis suggests that the intervention program has strong spillover effects through the students' networks.
By contrast, $\mathbb{H}^{\mathtt{b}}$ and $\mathbb{H}^{\mathtt{c}}$ are not rejected even under the significance level of 10\%.
Thus, we might conclude that the presence of at least one seed student in the student's neighborhood, rather than the number of treated friends, can reasonably explain the social interactions in the students' anti-conflict activities.

\begin{table}
	\caption{Empirical results: anti-conflict education program}
	\label{tab:paluck}
	\small
	\begin{center}
		\begin{tabular}[t]{rrrrrrrrrr}
			\toprule
			\multicolumn{2}{c}{ } & \multicolumn{3}{c}{$p$-values} & \multicolumn{3}{c}{Simes' correction} & \multicolumn{2}{c}{ } \\
			\cmidrule(l{3pt}r{3pt}){3-5} \cmidrule(l{3pt}r{3pt}){6-8}
			\multicolumn{1}{r}{$\mathcal{S}$} & \multicolumn{1}{c}{$\kappa$} & \multicolumn{1}{c}{KW} & \multicolumn{1}{c}{ACD} & \multicolumn{1}{c}{OLS} & \multicolumn{1}{c}{10\%} & \multicolumn{1}{c}{5\%} & \multicolumn{1}{c}{1\%} & \multicolumn{1}{c}{$|\mathcal{S}|$} & \multicolumn{1}{c}{$R$} \\
			\midrule
			\addlinespace[1em]
			\multicolumn{10}{l}{\underline{Testing for $\mathbb{H}^{\mathtt{a}}$}}\\
			\hspace{2.5cm} 3-net & 2 & 0.012 & 0.012 & 0.005 & \checkmark & \checkmark &  & 774 & 100,000 \\
			\hspace{2.5cm} 2-net & 2 & 0.025 & 0.025 & 0.012 & \checkmark & \checkmark &  & 966 & 100,000 \\
			\addlinespace[1em]
			\multicolumn{10}{l}{\underline{Testing for $\mathbb{H}^{\mathtt{b}}$}}\\
			\hspace{2.5cm} 3-net & 3 & 0.257 & 0.256 & 0.257 &  &  &  & 681 & 100,000 \\
			\hspace{2.5cm} 2-net & 3 & 0.126 & 0.144 & 0.126 &  &  &  & 817 & 100,000 \\
			\addlinespace[1em]
			\multicolumn{10}{l}{\underline{Testing for $\mathbb{H}^{\mathtt{c}}$}}\\
			\hspace{2.5cm} 3-net & 2 & 0.442 & 0.387 & 0.473 &  &  &  & 284 & 100,000 \\
			\hspace{2.5cm} 2-net & 2 & 0.681 & 0.639 & 0.777 &  &  &  & 291 & 100,000 \\
			\bottomrule
		\end{tabular}
	\end{center}
\end{table}

\section{Conclusion}\label{sec:conclusion}

We developed a novel randomization testing approach for the specification of general exposure mappings in treatment effect models with interference.
Based on the concept of coarseness of exposure mappings, our proposed approach has a fairly broad empirical applicability and enables us to construct model-free test statistics with a good power property.
As empirical illustrations, we have revisited two existing social network experiments in the literature: one is the data on farmers' insurance adoption studied in \cite{cai2015social} and the other is the data on anti-conflict education programs studied in \cite{paluck2016changing}.
From the results of the experiments on both datasets, we found that the exposure mapping $E_i(\bm{z}) = \left( z_i, \max_{j \in \mathcal{P}_i} z_j \right)$ has a certain capability to account for the spillover effects.

\section*{Acknowledgments}

We thank Jing Cai for kindly instructing us on how to recover the randomization strata from the replication data of \citet{cai2015social}.
We also thank Sukjin Han, Marc Henry, Ryo Okui, Kohei Yata, and conference and seminar participants at IAAE 2023, Keio University, Kwansei Gakuin University, The University of Tokyo, and Osaka University for their helpful comments and discussions.
This work was supported by JSPS KAKENHI grant numbers 19H01473 and 20K01597.
The datasets used in the empirical illustrations are available from the Interuniversity Consortium for Political and Social Research (\citealp{cai2019replication}; \citealp{paluck2020changing}).

\clearpage

\bibliography{references.bib}

\end{document}